\documentclass{emulateapj}
\usepackage{graphicx}

\newcommand{\RA}[3]{\mbox{R.A.}={#1}$^{{\rm h}}${#2}$^{{\rm m}}${#3}$^{{\rm s}}$}
\newcommand{\decl}[3]{\mbox{Dec.}={#1}$^{\circ}${#2}\arcmin{#3}\arcsec}
\newcommand{\arcs}{\arcsec}
\newcommand{\degree}{$^{\circ}$}

\newcommand{\cm}[1]{~cm$^{#1}$}

\newcommand{\e}[1]{10$^{#1}$}
\newcommand{\ee}[1]{$\times$10$^{#1}$}
\newcommand{\ergs}{~erg\,cm$^{-2}$\,s$^{-1}$}

\newcommand{\msun}{M$_{\odot}$}

\shorttitle{XMM-Newton observations of GW150914}
\shortauthors{Troja et al.}

\begin{document}
\title{XMM-Newton Slew Survey observations of the gravitational wave event GW150914}

\author{E. ~Troja\altaffilmark{1,2},
A.~M.~Read\altaffilmark{3},
A.~Tiengo\altaffilmark{4,5,6},
R.~Salvaterra\altaffilmark{5}
}

\altaffiltext{1}{NASA Goddard Space Flight Center, 8800 Greenbelt Rd, Greenbelt, MD 20771, USA.}
\altaffiltext{2}{Department of Physics and Astronomy, University of Maryland, College Park, MD 20742-4111, USA.}		
\altaffiltext{3}{Department of Physics and Astronomy, Leicester University, Leicester LE1 7RH, UK. }
\altaffiltext{4}{Istituto Universitario di Studi Superiori, piazza della Vittoria 15, I-27100 Pavia, Italy.}
\altaffiltext{5}{Istituto di Astrofisica Spaziale e Fisica Cosmica Milano, INAF, via E. Bassini 15, I-20133 Milano, Italy.}
\altaffiltext{6}{Istituto Nazionale di Fisica Nucleare, Sezione di Pavia, via A. Bassi 6, I-27100 Pavia, Italy.}

\begin{abstract}
The detection of the first gravitational wave (GW) transient GW150914 prompted an extensive campaign of follow-up observations at all wavelengths. 
Although no dedicated {\it XMM-Newton} observations have been performed, the satellite passed through the GW150914 error region during normal operations. 
Here we report the analysis of the data taken during these satellite slews performed two hours and two weeks after the GW event.  Our data
cover 1.1 deg$^2$ and 4.8 deg$^2$ of the final GW localization region. 
No X-ray counterpart to GW150914 is found down to a sensitivity of $6\times 10^{-13}$ ergs cm$^{-2}$ s$^{-1}$ in the 0.2--2 keV band. 
Nevertheless, these observations show the great potential of {\it XMM-Newton} slew observations for the search of the electromagnetic counterparts of GW events. 
A series of adjacent slews performed in response to a GW trigger would take $\lesssim$1.5 days to cover most of the typical GW credible region. 
We discuss this scenario and its prospects
for detecting the X-ray counterpart of future GW detections.

\end{abstract}

\keywords{gravitational waves; gamma-ray bursts; X-rays}

\section{Introduction}\label{sec:intro}

On 2015 September 14 at 09:50:45 UTC, the advanced Laser Interferometer Gravitational-Wave Observatory (LIGO; \citealt{aligo}) triggered and detected its first gravitational wave event,
dubbed GW150914 \citep{ligodet16}.  This was a revolutionary discovery, which confirmed the predictions of general relativity \citep{ligogw16} 
and opened a new window into the study of our universe.
The refined LIGO analysis yields a robust detection \citep{ligofar16}, and shows that the observed waveform
is consistent with the merger of  a binary black hole (BBH) system with relatively ``heavy'' masses of $\approx$36\,\msun\ and $\approx$29\,\msun, thus providing the first observational evidence that these astrophysical systems exist \citep{belc10, ligoastro16}. 

Despite its crude localization ($\approx$590 deg$^2$), GW150914 enjoyed an extensive observing campaign across the electromagnetic spectrum \citep{ligoem16}. 
Although no visible counterpart was expected from a BBH merger, \citet{valerie16} reported the presence of a weak, short duration ($\sim$1~s) gamma-ray excess, starting 0.4~s after the GW trigger and coming from a sky area consistent with the GW localization.  This event was not detected during the more sensitive INTEGRAL observations \citep{integral}, thus its nature as well as its association with GW150914 remain uncertain. 
Nonetheless, several new scenarios were proposed in order to explore the possible electromagnetic signatures of BBH merger, including short gamma-ray bursts and their accompanying afterglows \citep[e.g.][]{perna16,murase16}.  In this respect, X-ray observations of GW events represent a promising route towards the 
discovery of their electromagnetic counterparts. The X-ray window offer several advantages when compared to other energy bands: 
a) most ($\sim$80\%) short GRBs have an X-ray afterglow detection \citep{davanzo14}, compared to only 30\% detected in the optical band \citep{kann11}, showing that X-ray observations are far more efficient in detecting these events;
b) space-based X-ray observations are not subject to atmospheric constraints, that often hamper or degrade ground-based searches;
c) the number of X-ray candidates expected within the LIGO localizations is significantly lower than in deep optical surveys.

The major obstacle for X-ray observatories is to rapidly observe the large GW error regions with an adequate sensitivity. 
In the case of GW1509014,  the entire LIGO error region was scanned by MAXI with a shallow sensitivity of $\approx$\e{-9}\ergs\ in the 2-20 keV energy band \citep{ligoem16}. 
Deeper Target of Opportunity (ToO) observations of the field were performed by the {\it Swift} X-ray Telescope, which covered an area of 5 deg$^2$ with a sensitivity of 6\ee{-13} - 6\ee{-12}\,\ergs \citep{evans16}. In this work, we report the serendipitous {\it XMM-Newton}  \citep{jansen01} slew survey observations of the LIGO localization. 
In section~\ref{sec:obs} we describe our observations, in section~\ref{sec:res} we present our results, and in section~\ref{sec:slew}  we propose a novel observational strategy in order to maximize the impact of {\it XMM-Newton} in the era of gravitational wave astronomy. 

\begin{figure*}
\begin{center}
\includegraphics[width=12cm,angle=0]{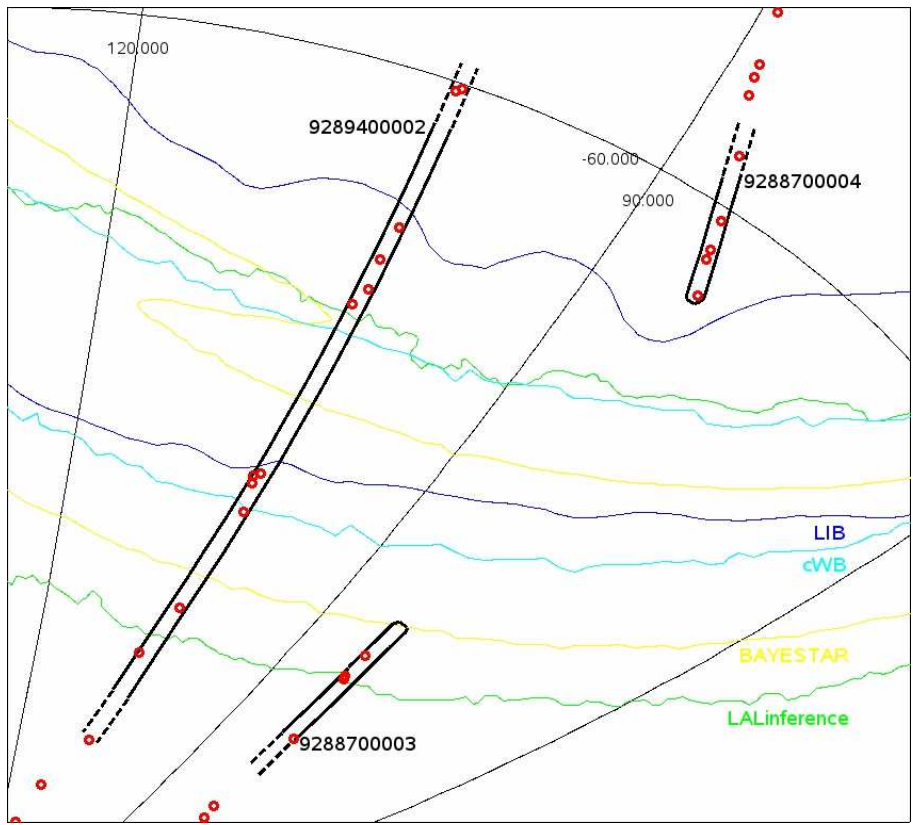}
\end{center}
\caption{A map of the {\it XMM-Newton} slews and slew sources from the two weeks after the GW150914
  event. Slew sources are shown as the small
  circles. Slews are shown as the thin black strips (9288700003 extends to the
south, 928870004 to the north and 9289400002 in both directions). The LIGO
 90\% confidence regions from the four different pipelines - LALInference,
BAYESTAR, cWB (sky) and LIB - are shown. No other
slews in the two week period after the GW150914 event lie anywhere close to
the LIGO localization regions.}
\label{fig1}
\end{figure*}

\section{Observations}\label{sec:obs}

The {\it XMM-Newton} Slew Survey is based solely on data from the European Photon Imaging Camera (EPIC) pn camera \citep{struder01}.
The in-orbit slew speed of 90$^{\circ}$ per hour results in an exposure time between 1--11~s.
The soft band (0.2$-$2 keV) sensitivity limit of {\it XMM-Newton} slews is
6$\times10^{-13}$\,ergs cm$^{-2}$ s$^{-1}$, close to that of the ROSAT
All-Sky Survey (RASS; \citealt{voges99}). In the hard (2$-$12 keV) band
the slew data goes significantly deeper (4$\times10^{-12}$\,ergs cm$^{-2}$
s$^{-1}$) than previous surveys. For details on the
construction and characteristics of the {\it XMM-Newton} slew survey
catalogue, see \citet{saxton08}.

At the time of the GW150914 event, {\it XMM-Newton} was performing a pointed
observation towards the direction of \RA{18}{23}{14.15}, \decl{-01}{27}{38.3}, far from the
LIGO localization region of GW150914. Shortly afterwards, the satellite
slewed and crossed by chance the LIGO 90\% error region.  Many slews
were subsequently made across the sky as the spacecraft moved between its
scheduled pointing positions.

Here we analyze the {\it XMM-Newton} slews made in the two weeks directly after the
GW150914 event. Six slews were made during that time, but only three intercept or come close the 
LIGO localization maps, as listed in Table~\ref{tab1}.
The data reduction and analysis used is in many ways similar to the standard
slew analysis \citep{saxton08}. However a number of
improvements to {\it XMM-Newton} slew data analysis have been made, and were used in
the present analysis, and these will be described in a future paper (Read et al, in prep).

\begin{table}[!h]
\caption{Log of XMM-Newton Slew Observations.} 
\begin{center}       
\begin{tabular}{ccccc} 
\hline
\hline
OBSID & Start Time & T-T$_{GW}$ & \multicolumn{2}{c}{LIGO Coverage} \\
  & (UT) &   &  All &  LALInf \\
  &      &   &  (deg$^2$)&(deg$^2$) \\
\hline
9288700003 & 2015-09-14 11:55:03 & 2.1 hrs & 1.1 & 1.1  \\
9288700004 & 2015-09-15 01:34:53 & 0.66 d & -- &  --   \\
9289400002 & 2015-09-28 23:28:39 & 14.6 d & 6.5 & 4.8  \\
\hline
\end{tabular}
\end{center}
\label{tab1}
\end{table} 

\begin{table*}
\caption{XMM-Newton slew sources detected within the total LIGO localization region.} 
\begin{center}       
\begin{tabular}{ccccccccccc} 
\hline
\hline
Name & RA    & Dec     & \multicolumn{3}{c}{Source Counts} & \multicolumn{3}{c}{Detection Likelihood} & Quality & Catalog ID \\
           & (deg) & (deg)  & 0.2-12 & 2-12 & 0.2-2                      & 0.2-12 & 2-12 & 0.2-2  & \\
  & & & keV & keV & keV & keV & keV & keV & \\
\hline
\multicolumn{11}{c}{OBSID: 928870003}\\
\hline
XSS~J052742.8-763133  & 81.92836  & -76.52586  & 3.9 &  --	& 3.9 &10.5 &  --	  & 10.4  	& 2 &  1SWXRT J052742.4-763128 \\
XSS~J052914.5-762522 & 82.31060  & -76.42279 &15.2 & 4.4 & 9.9 &48.4 & 10.2      & 25.2 	& 1 &  1RXS~J052918.2-762514 \\
XSS~J052952.5-753757 & 82.46876  & -75.63253 & 3.6 & --	& --    & 8.8 &  --	  &  	  -- 	& 3 & WISE~J052952.46-753805.9 \\
\hline
\multicolumn{11}{c}{OBSID: 928940002}\\
\hline

XSS~J064838.0-641619 & 102.15847 & -64.27199 & 3.7 & --	& 3.8 & 9.7 &  --	 & 10.3 	& 1 & 1RXS~J064837.2-641624	 \\
XSS~J064903.4-661054 & 102.26428 & -66.18164 & 5.3 & --	& 5.5 &13.0 & -- 	 & 15.1 	& 2 & WISE~J064903.95-661045.2\\
XSS~J064943.8-651636 & 102.43274 & -65.27666 & 3.2 & --	&  --   & 9.0 &  --	 &  --	   	& 3 & WISE~J064943.68-651638.7\\  
XSS~J065106.2-664320 & 102.77593 & -66.72232 & 3.8 & --	&  --   &11.0 &  --	 &  --	   	& 4 & -- \\ 
XSS~J065226.0-732221 & 103.10855 & -73.37250 & 3.9 & --	&  --   &10.1 &  --	 &  --	   	& 4 & -- \\  
XSS~J065327.3-720907 & 103.36414 & -72.15190 & 3.7 & --	&  --   & 9.5 &  --	 &  --	   	& 2 & WISE~J065327.99-720903.2\\  
XSS~J065436.5-722926 & 103.65240 & -72.49045 & --  & --	& 3.8   & --  &  --	 &  15.8	& 1 & 1RXS~J065433.3-722928\\  
XSS~J065520.8-721745 & 103.83708 & -72.29580 & 5.5 & --	&   --  & 8.3 & -- 	 &  --	   	& 3 & WISE~J065521.78-721738.0\\
XSS~J065720.5-764311 & 104.33549 & -76.71979 & 3.5 & --	&  --   &10.1 & -- 	 &  --	   	& 2 &  ATPMN~J065720.9-764309\\
\hline
\end{tabular}
\end{center}
\label{tab2}
\end{table*} 


\section{Results}\label{sec:res}

A detailed comparison of the slew coverage with the localization regions
from the four different LIGO pipelines - LALInference, BAYESTAR, cWB (sky) and LIB
-  is shown in Fig.\ref{fig1}. Since GW150914 is a compact binary coalescence (CBC) 
event, it is considered that the LALInference map is the most accurate,
authoritative, and final localization for this event \citep{ligoem16}. We therefore prioritize
the LALInference map, but still consider all four maps together as a wider total
region. 
We calculate that slew 9288700003 covers $\approx$1.1\,deg$^2$ of the
LALInf (and of the total) region,  and slew 9289400002 covers $\approx$4.8\,deg$^2$ of the LALInf region (and $\approx$6.5\,deg$^2$ of the total region). 
Slew 9288700004 marginally overlaps with the LIGO localization and only covers an area outside the 90\% confidence region.
These values are tabulated in Table~\ref{tab1}.

Source detection was performed via usage of a semi-standard \texttt{eboxdetect}
(local) + \texttt{esplinemap} + \texttt{eboxdetect} (map) + \texttt{emldetect} method, and performed on
a single image containing the single events (pattern=0) in the 0.2--ˆ'0.5
keV band, plus single and double events (pattern=0--ˆ'4) in the 0.5--12 keV band. 
Source-detection is performed separately in three
energy bands; a total band (0.2-12\,keV), a hard band (2-12\,keV) and a soft
band (0.2-2\,keV). 
The source position and background maps are computed separately in each band, and this may result in small differences
between the total band and the soft+hard band.
The sources detected in the three bands were then combined
to produce an initial catalogue. Twelve sources were detected in the slew
areas coincident with the (total) LIGO region, and X-ray parameters for these
are given in Table~\ref{tab2}. The sources are also indicated (together with
sources detected outside of the LIGO region) in Figure~\ref{fig1}. 
Each detection was assigned a quality flag, from 1 (likely real) to 4~(likely spurious). 
The detection likelihood (DET\_ML) was computed by the \texttt{emldetect} task and is defined as  $DET\_ML = - ln~P$, where $P$ is the
probability the detection is spurious due to a Poissonian fluctuation \citep{watson09}.

 \citet{saxton08} quote the
astrometric uncertainty of slew sources to be about 8\arcs (68\%
confidence error radius), and  \citet{warwick12} quote
a 90\% error circle radius for their slew sample of 10\arcs. 
Searches for multiwavelength counterparts made use of the facilities at Vizier, Simbad and NED.
Specific cross-correlations were made with WISE and with 2MASS (as in \citealt{warwick12}), 
and in the case of ROSAT, extra allowances were made to account for the uncertainty in the ROSAT positions.
The results are reported in Table~\ref{tab2} (last column).
Apart from XSS~J065106.2-664320 and XSS~J065226.0-732221, the remaining ten sources
all have WISE counterparts within~10\arcs. The chance of a random positional coincidence with a WISE source is only 20\%.
Among them, four correspond to catalogued X-ray sources. 
One source, XSS~J064838.0-641619, is coincident with the nearby elliptical galaxy NGC~2305 at a distance of  48~Mpc. 
However, it is detected at a flux level consistent with the ROSAT observations. 
Based on Table~2, we conclude that no new X-ray source was detected in our observation.

\begin{figure*}[!th]
\begin{center}
\includegraphics[scale=0.4,angle=0]{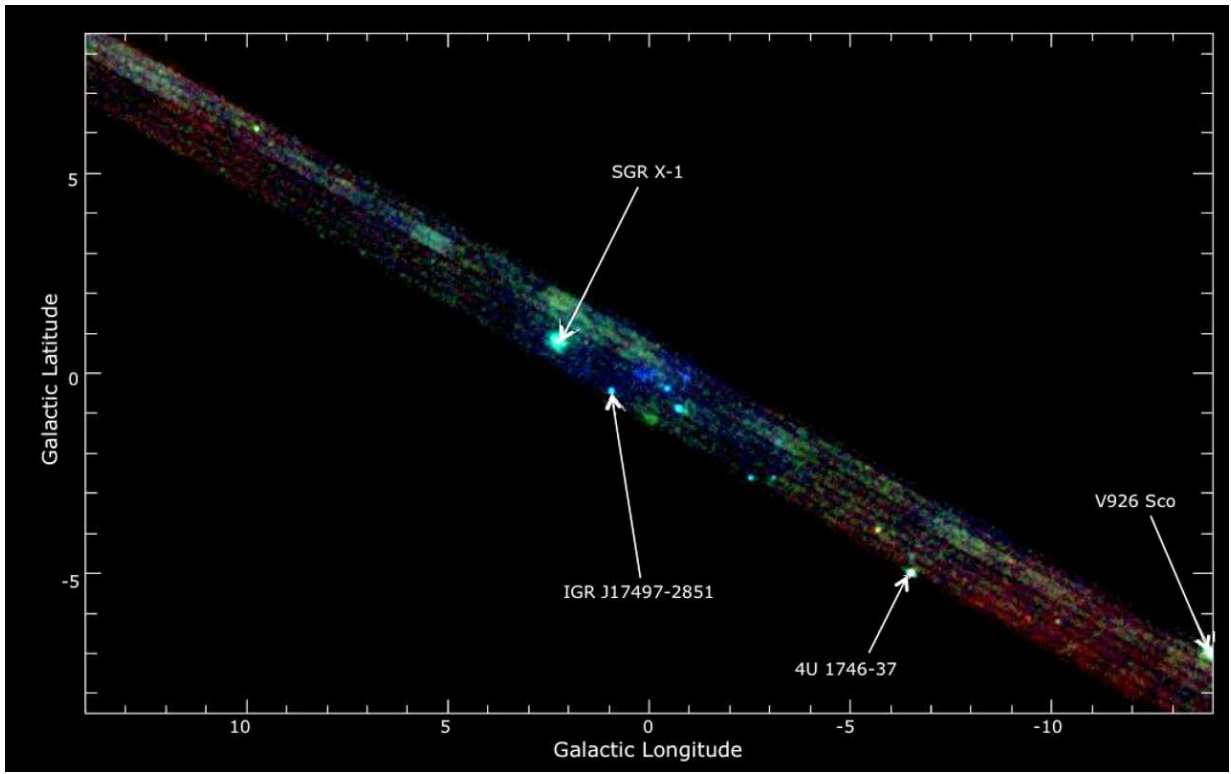}
\end{center}
\caption{Image of the Galactic Center obtained by slewing {\it XMM-Newton} during revolution 1242. The image covers a 40$\times$3.5~deg$^2$ area, comparable to LIGO localizations. 
This test was performed at a reduced slew speed of 30\degree~per hour,  which is not required in our case. 
We propose to perform slew observations of future GW triggers at a standard slew speed of 90\degree~per hour. 
This would allow us to cover a 20~deg$^{2}$ area in $\approx$30 ks.  }
\label{fig:sss}
\end{figure*}

\section{Strategy}\label{sec:slew}

{\it XMM-Newton} mainly operates in pointing mode, with a minimum exposure of 5~ks per observation.  This time-constraint makes highly impractical 
to rapidly tile large areas of the sky or to follow-up a large number of candidates.
Slew observations are executed only between two pointed observations. Indeed, the 90\% error region of GW150914 was partially covered by {\it XMM-Newton} slews only by chance. 
Nevertheless, the observations presented in the previous section covered 6 deg$^2$ (8 deg$^2$) of the LALInf (total) credible region with a sensitivity comparable to the {\it Swift}/XRT follow-up \citep{evans16}.
Here we explore the possibility to perform {\it XMM-Newton} Target of Opportunity Slew Surveys (ToOSS) in response to future GW triggers.

Given the 14$^{\prime}$ radius EPIC-pn field of view and the slew speed of 90$^{\circ}/$hr, the 590 deg$^2$ area of the GW150914 error region could be in principle covered in about 14 hours. However, overhead times could significantly increase the total time required to perform such a large area survey.
Their impact can be evaluated on the basis of 
a special test for a large area slew survey that was performed at a reduced speed in September 2006 (revolution 1242).
This slow slew survey covered a 140 deg$^2$ sky region (a rectangle of $40^{\circ}\times3.5^{\circ}$; see Figure~\ref{fig:sss}) within a single satellite orbit 
($\sim$1.5 days of scientific observing time) with 16 partially overlapping slews at a reduced speed of 30$^{\circ}/$hr. During this test, the time overhead between two adjacent slews was $\sim$3 ks.  
A similar survey at the normal slew speed of 90$^{\circ}/$hr would take only $\sim$20 hours and an area larger than 40\% of the GW150914 error region could be covered within a single {\it XMM-Newton} orbit. Without requiring an overlap of the slews, the area surveyed in an orbit would increase to more than 70\% of the GW150914 error region.
Furthermore, in future runs the localization accuracy of GW transients is expected to improve thanks to the increased sensitivity, larger bandwidth and the addition of other detectors to the GW network.
A 50\% reduction in the error region is already expected for the O2 run (2016-2017), with 14\% of the detections being localized within 20~deg$^2$ \citep{ligoloc16}, an area which could be covered by slews in  $\approx$30~ks. 
By 2019 this fraction is expected to increase up to $\approx$30\% \citep{ligoloc16}.

The actual observing efficiency depends also on the sky position of the GW trigger, which determines the slew length, its direction and the sky regions that cannot be observed due to the satellite visibility constraints. About 1/6 of the sky is accessible by {\it XMM-Newton}  at any given time, and some slewing directions may require more complex satellite operations and produce a worse astrometric accuracy, 
possibly causing an imperfect coverage of the region to be surveyed. 
Nonetheless, even if the whole error region cannot be covered due to visibility and slewing constraints, the proposed strategy will be an efficient way to search a significant fraction of a large sky area for an X-ray counterpart. 

The typical {\it XMM-Newton}  response time to a ToO trigger is $\lesssim$10--12 hours and, in the best case, it can be as small as 4--5 hours \citep[e.g.][]{xmmgcn}.
As we discuss later on, a very fast turn-around would be preferable in some exceptional cases (e.g. a joint GRB-GW detection or a well-localized GW source), 
and a slower reaction time is instead preferred to search for off-axis afterglow emission \citep{granot02,vaneerten11}.
In order to maximize the scientific return,  the data should be analyzed in ``real time" to identify  and localize possible candidates 
with sufficient accuracy to enable further follow-up campaigns. At the moment a fast automatic analysis is performed on slew data, but it requires orbital data that become available with a delay of 1--2 weeks. 
Without the orbital data, the slew source positions within a few arcmin could be obtained directly from raw slew data, relying on the nominal slewing direction and the relative position of known bright X-ray sources. 
This would make it possible to reduce the processing time to few hours.

\begin{figure*}[!t]
\begin{center}
\includegraphics[scale=0.33,angle=0]{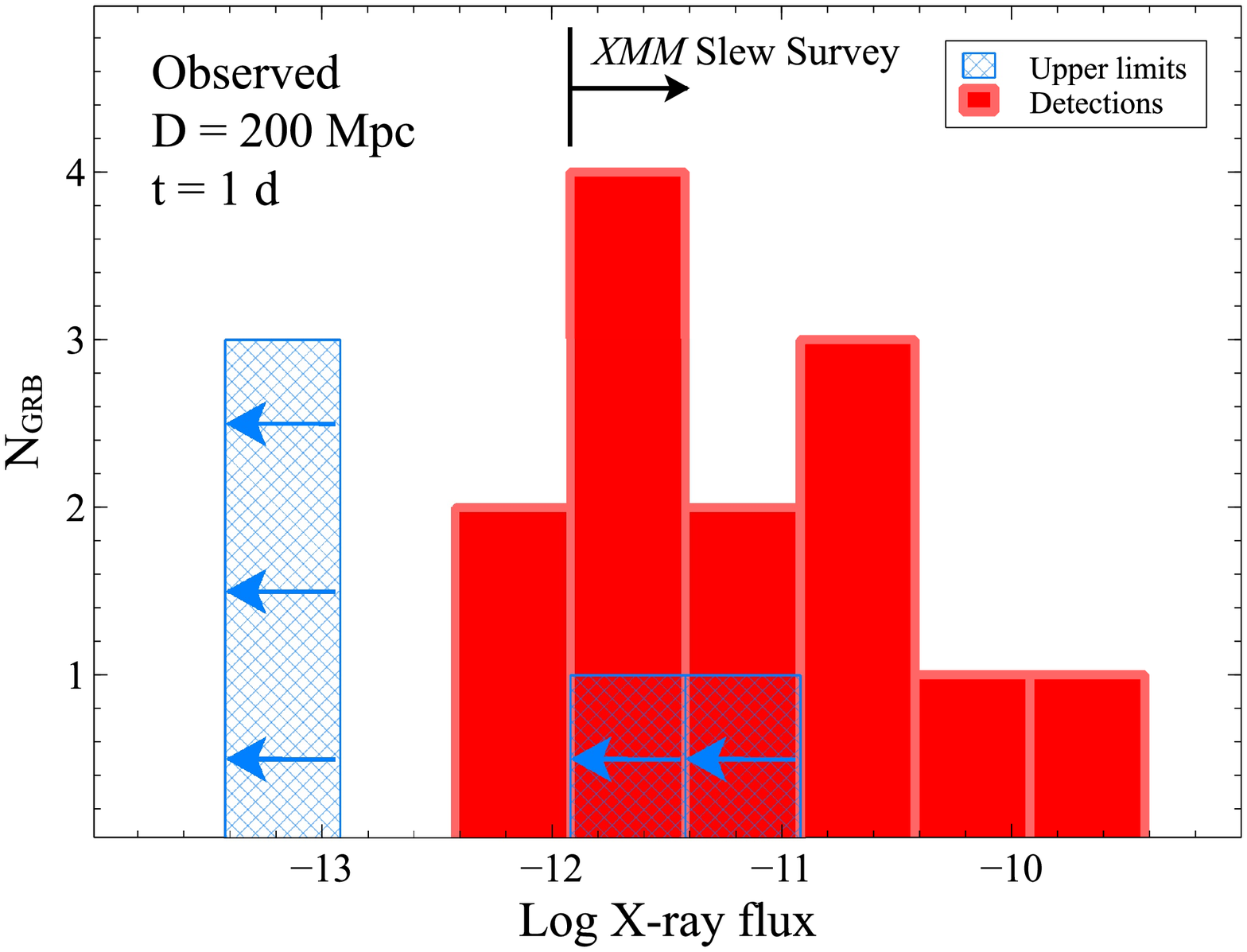}
\hspace{0.4cm}
\includegraphics[scale=0.33,angle=0]{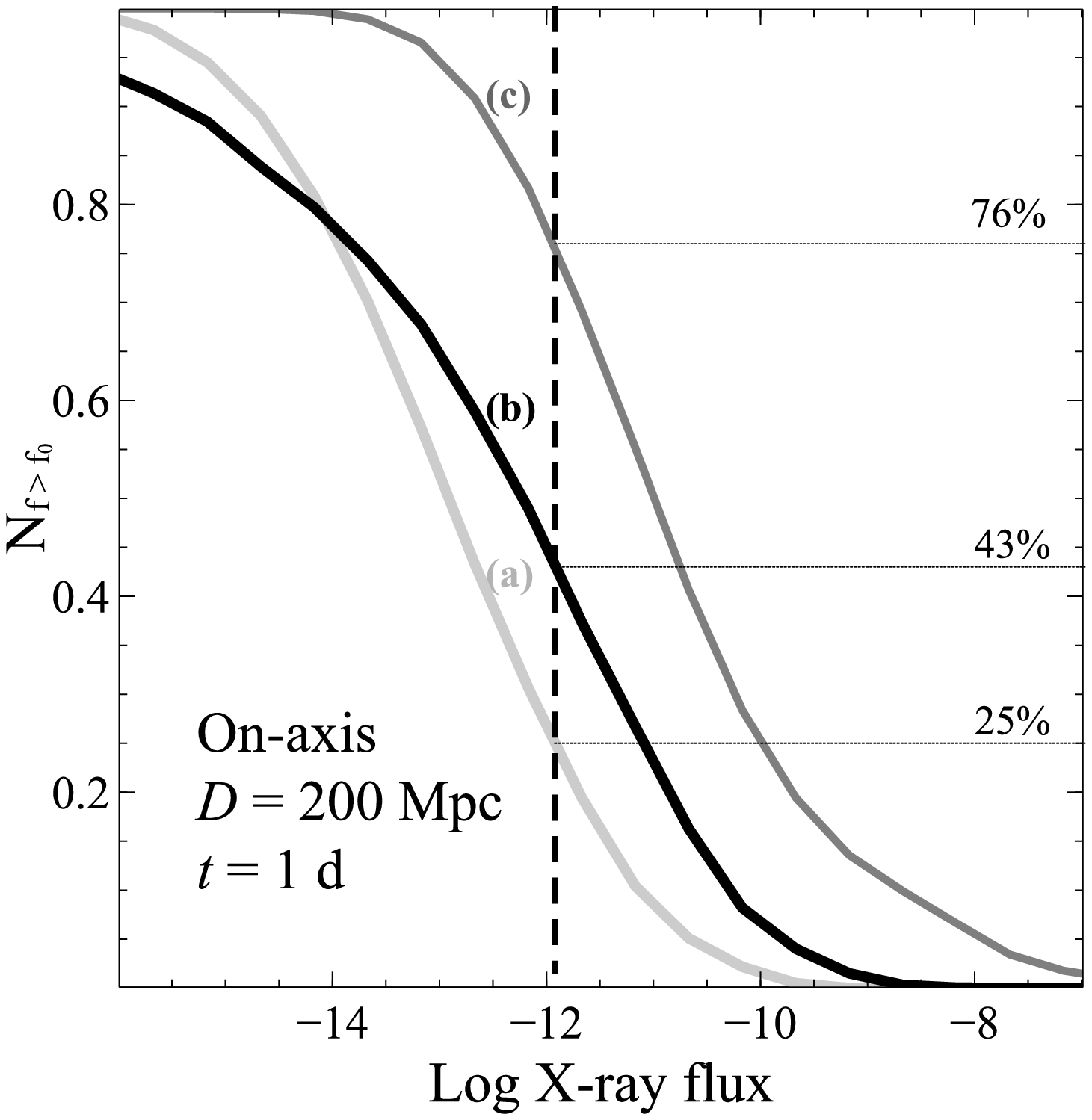}
\hspace{0.4cm}
\includegraphics[scale=0.33,angle=0]{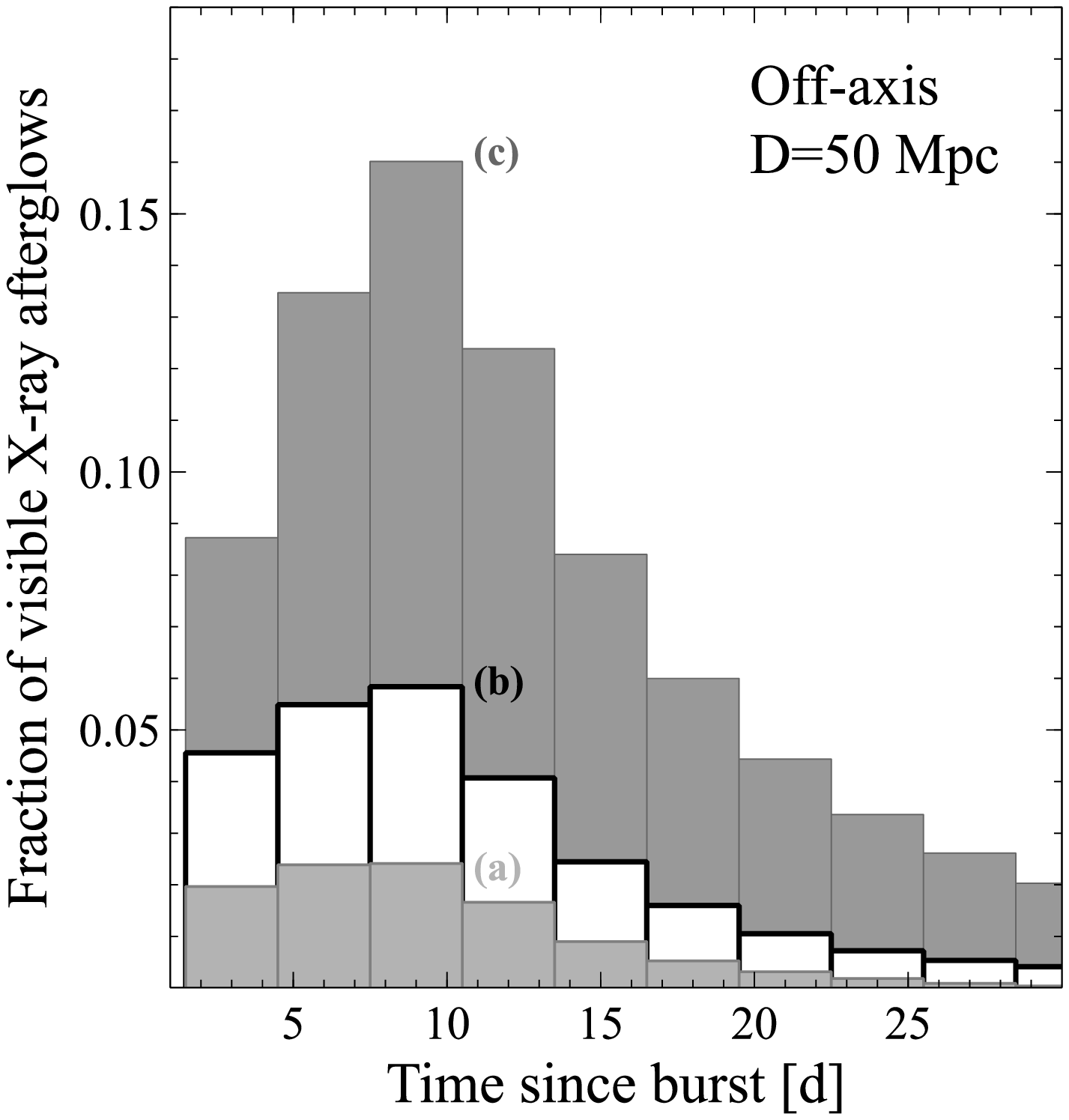}
\end{center}
\caption{\textit{Left panel:} X-ray flux distribution for short GRB afterglows with known distance. 
Fluxes were calculated in the 0.2-12 keV energy band at a time $t$=1~d after the burst onset. Values were then redshifted at a common distance $D$=200~Mpc.
\textit{Middle panel:}  Cumulative flux distribution for simulated on-axis X-ray afterglows at $D$=200~Mpc, and at a time $t$=1~d post-burst. The input parameters used for the simulations are described in the text. 
The vertical dashed line indicates the {\it XMM-Newton} slew sensitivity in the 0.2-12~keV energy band. Depending on the details of the GRB explosion the
fraction of visible afterglows ranges between $\approx$25\% and $\approx$75\%. 
\textit{Right panel:}  Fraction of visible off-axis X-ray afterglows as a function of time. We assumed a distance of $D$=50~Mpc,
and the same input parameters of on-axis simulations. 
The peak ranges between $\approx$3\% and $\approx$15\% at $t$ $\approx$8 days post-burst.
}
\vspace{0.22cm}
\label{fig2}
\end{figure*}

In Figure~\ref{fig2} we estimate the detectability of short GRB afterglows during an {\it XMM-Newton} slew observation.  
First, we considered the known sample of {\it Swift} short GRBs with measured redshift \citep[e.g.][]{davanzo14}. 
The left panel shows their X-ray fluxes distribution at $t$=1~d and at a distance $D$=200 Mpc, which is the horizon distance for NS-NS mergers at the LIGO design sensitivity. 
Most of them ($>$60\%) lie above the {\it XMM-Newton} slew survey detection threshold, $f_X$=1.2\ee{-12}\,\ergs (0.2--12~keV).
However, one has to consider that the observed sample is shaped by complex observational biases, which tend to favor the detection and localization of the brightest events. 
We therefore simulated a sample of GRB afterglows following standard prescriptions for a spherical fireball \citep{gs02}, and post-jet-break scalings from \citet{sari99}. 
The explosion properties of short GRBs are not well constrained, and we made an educated guess about the values and distributions of afterglow parameters.
We assumed an electron spectral index $p$$\approx$2.3, and simulated three populations of explosions with the following parameters: 

a) a total energy release $E$ uniformly distributed between \e{48}\,erg and \e{51}\,erg, 
a medium density $n$ between 0.001 and 1 \cm{-3}, 
an electron energy fraction $\epsilon_e$  between \e{-3} and 0.1, 
a magnetic energy fraction $\epsilon_B$ between \e{-4}  and \e{-2}, 
and an opening angle $\theta_j$ between~1$^{\circ}$~and~40$^{\circ}$.

b) a total energy release with average $<$$E$$>$=\e{49}\,erg and gaussian dispersion $\sigma_{E}$=1 dex, 
$n$ between 0.001 and 1 \cm{-3}, 
$<$$\epsilon_e$$>$=0.1 with gaussian dispersion $\sigma_{e}$=0.5 dex, 
$\epsilon_B$ between \e{-4}  and \e{-2}, 
and $<$$\theta_j$$>$=10$^{\circ}$ with gaussian dispersion $\sigma_{\theta}$=0.5 dex. 

c) $<$$E$$>$=\e{50}\,erg with gaussian dispersion $\sigma_{E}$=1 dex, 
$<$$n$$>$=0.1 with $\sigma_{n}$=1 dex, 
$<$$\epsilon_e$$>$=0.1 with $\sigma_{e}$=0.5 dex, 
$<$$\epsilon_B$$>$=0.1 with $\sigma_{B}$=0.5 dex, 
and $<$$\theta_j$$>$=10$^{\circ}$ with  $\sigma_{\theta}$=0.5 dex.

For each set of parameters, we simulated 340,000 afterglows and calculated their X-ray fluxes in the 0.2-12 keV at $t$=1~day. 
We assumed an intrinsic absorption N$_{H,i}$=\e{21}\cm{-2}, and a flat galactic column density N$_{H,g}$=3\ee{20}\cm{-2}.
The results are shown in Figure~\ref{fig2} (middle panel), where we report the cumulative flux distribution.
For the most optimistic case (c) the fraction of visible X-ray afterglows is nearly 80\%, and for the less optimistic
scenario (a) is still 25\%.
These estimates should be obviously taken with a grain of salt as they simplify a complex and poorly constrained phenomenon.
The predicted fluxes, calculated for on-axis observers, are significantly fainter if the jet is seen far off-axis ($\theta_{obs}$$>$$\theta_j$).  
Viewing angle effects could therefore affect our calculations, decreasing the number of visible X-ray afterglows by a factor $\sim$3 to over 100,
depending on the (unknown) beaming distribution of short GRB jets.
This may be alleviated by the fact that the source inclination impacts not only the afterglow flux, but also the strength of the emitted gravitational radiation.
 Face-on binaries, i.e. with their rotation axis pointing toward us, are more likely to be detected by a factor of~3.4~\citep{schutz11}. 
It is plausible that the GRB jet will form along the same axis, thus suggesting that on-axis short GRBs should also be 
stronger GW sources. 
A joint GW-GRB detection would give us a high level of confidence to search for an on-axis X-ray afterglow. However, 
even in this case, a standard follow-up strategy could not be effective. Given its large field of view,
the {\it Fermi} Gamma-ray Burst Monitor is most likely to detect such a joint event, thus providing only a coarse localization \citep[e.g.][]{valerie16} comparable to the LIGO one. 

In absence of a GRB trigger, the most likely counterpart would be an orphan afterglow, whose emission is much fainter, and therefore detectable within a much smaller volume. By adopting the simple analytical model of \citet{granot02} with the same explosion parameters listed above, 
and folding in the probability distribution of \citet{schutz11} for the binary orientation, 
we derive in Figure~\ref{fig2}~(right panel) the prospects for detecting such off-axis emission with {\it XMM-Newton}. 
For a distance of 50~Mpc, the fraction of detectable events ranges between 3\% and 15\%, much smaller than for an on-axis explosion. 
However,  Figure~\ref{fig2} also shows that the optimal reaction time is slow, peaking between 7 and 15 days after the burst.

Finally, we consider that the past decade of afterglow studies has unveiled the presence of new emission components in addition to the standard forward shock emission described in \citet{gs02}. Of particular interest is the emission from a stable millisecond magnetar, which could power a bright and nearly isotropic X-ray transient.
\citet{zhang13} estimated a bright and persistent X-ray emission from a magnetar-driven relativistic wind lasting for several hours after the burst onset. 
\citet{metzger14} considered instead the emission from the remnant pulsar wind nebula, and estimated a peak X-ray luminosity \e{43}-\e{44}~erg\,s$^{-1}$ on a timescale $\lesssim$1 day.  At a distance of 200 Mpc, this corresponds to a flux 2\ee{-12}-2\ee{-11}\,\ergs, above the {\it XMM-Newton} slew survey sensitivity.  
In case of a particularly well-localized GW event ($\approx$20 deg$^2$), we argue that a rapid observation of the GW field would still be extremely valuable in order to search for these possible X-ray counterparts. Within this field we expect to detect on average 10 X-ray candidates with high-significance, which can be easily targeted for subsequent follow-up with {\it Swift} or optical/nIR ground-based facilities.

\section{Summary}\label{sec:end}

We reported the {\it XMM-Newton} observations of the field of GW150914, the first LIGO detection. 
Although no X-ray counterpart was found, these serendipitous observations show the great potential of {\it XMM-Newton} slews 
to search for electromagnetic counterparts of GWs. A single slew of only 7 minutes covered 4.8 deg$^2$ of the LIGO region down
to a sensitivity of 6$\times$10$^{-13}$ erg cm$^{-2}$ s $^{-1}$ (0.2--2 keV). 
An observing strategy consisting of a series of adjacent slews (Figure~2)  could survey a large fraction of future
LIGO localizations within a single {\it XMM-Newton} orbit.
In order to maximize the chances of success without excessively impacting {\it XMM-Newton} operations, we suggest three possible responses to future GW triggers: a) a rapid ($\lesssim$1\,d) high-priority ToO of the GW region in case of a simultaneous GRB trigger with poor localization; b) a slower response ($\approx$7 days) ToO in case of a rare, nearby ($\lesssim$50~Mpc) GW transient; c) a rapid ToO in case of a
well-localized ($\approx$20~deg$^2$) GW event. The latter case could be assigned a medium-priority status, i.e. implemented only during working hours.
On the basis of the experience of the {\it XMM-Newton}  Slew Survey, we expect about 0.5 candidates per square degree, 
much less than the sources expected from deep optical surveys.  Multi-wavelength follow-up observations may be required 
in order to characterize and identify the selected sources. 
For a canonical fireball spectral index $\beta_{OX}$$\gtrsim$0.6,  the optical counterparts of {\it XMM-Newton} slew candidates are expected to have $r$$\lesssim$22~mag, 
which can be reached with small aperture ($\sim$1.5~m) telescopes in reasonably short ($\lesssim$10~min) exposures.

\acknowledgements
We thank Richard Saxton and the anonymous referee for useful and constructive comments which helped improve the manuscript.
This work is based on observations obtained with XMM-Newton, an ESA science mission with instruments and contributions directly funded by ESA Member States and NASA.
This research has made use of the SIMBAD database, operated at CDS, Strasbourg, France, and of the NASA/IPAC Extragalactic Database (NED) which is operated by the Jet Propulsion Laboratory, California Institute of Technology, under contract with the National Aeronautics and Space Administration.


\end{document}